\documentclass[nofootinbib,twocolumn,showpacs,preprintnumbers,pre,aps,superscriptaddress]{revtex4-1}
\usepackage[dvipdfmx]{graphicx}
\usepackage{bm}
\usepackage{amsmath}
\usepackage{amssymb}
\usepackage{color}

\usepackage{txfonts}
\DeclareMathAlphabet\mathbfcal{OMS}{cmsy}{b}{n}

\begin{document}
\title{Generalized Three-Sphere Microswimmers}

\author{Kento Yasuda}

\affiliation{
Research Institute for Mathematical Sciences, 
Kyoto University, Kyoto 606-8502, Japan}

\author{Yuto Hosaka}

\affiliation{
Max Planck Institute for Dynamics and Self-Organization (MPI DS), 
Am Fassberg 17, 37077 G\"{o}ttingen, Germany}

\author{Shigeyuki Komura}\email{komura@wiucas.ac.cn}

\affiliation{
Wenzhou Institute, University of Chinese Academy of Sciences, 
Wenzhou, Zhejiang 325001, China} 

\affiliation{
Oujiang Laboratory, Wenzhou, Zhejiang 325000, China}

\affiliation{
Department of Chemistry, Graduate School of Science,
Tokyo Metropolitan University, Tokyo 192-0397, Japan}

\begin{abstract}
Among several models for microswimmers, the three-sphere microswimmer proposed by Najafi 
and Golestanian captures the essential mechanism for the locomotion of a microswimmer in a viscous fluid.
Owing to its simplicity and flexibility, the original three-sphere model has been extended and 
generalized in various ways to discuss new swimming mechanisms of microswimmers.  
We shall provide a systematic and concise review of the various extensions of the three-sphere 
microswimmers that have been developed by the present authors.
In particular, we shall discuss the following seven cases; elastic, thermal, odd, autonomous 
three-sphere microswimmers; two interacting ones; and those in viscoelastic and structured fluids.
The well-known Purcell's scallop theorem can be generalized for stochastic three-sphere 
microswimmers and also for the locomotion in viscoelastic and structured fluids.  
\end{abstract}

\maketitle

\section{Introduction}

Microswimmers are tiny objects moving in fluids, such as sperm cells or motile bacteria, that swim in 
a fluid and are expected to be relevant to microfluidics and microsystems~\cite{Lauga09a}.
By transforming chemical energy into mechanical work, microswimmers change their shapes and 
move in viscous environments.
The fluid forces acting on the length scale of microswimmers are governed by the effect of viscous 
dissipation. 
According to Purcell's scallop theorem, reciprocal body motion cannot be used for locomotion 
in a Newtonian fluid~\cite{Purcell77,Ishimoto12}.
As one of the simplest models exhibiting non-reciprocal body motion, Najafi and Golestanian 
proposed a model of a three-sphere microswimmer~\cite{Golestanian04,Golestanian08}, in which three in-line 
spheres are linked by two arms of varying lengths. 
This model is suitable for analytical studies because the tensorial structure of the fluid motion can 
be neglected in its translational motion.
Later, such a three-sphere microswimmer has been experimentally realized~\cite{Leoni08,Grosjean16,Grosjean18}.

Owing to its simplicity and flexibility, the three-sphere model has been extended and generalized 
in different ways.  
For example, the two arms in the original model can be replaced by two elastic springs with 
time-dependent natural lengths~\cite{Pande15,Pande17,Yasuda17a,Ziegler19}.
Such an elastic three-sphere microswimmer can include the effects of thermal fluctuations acting 
on each sphere~\cite{Hosaka17,Yasuda21,Kobayashi23}. 
Then, Purcell's scallop theorem for a deterministic microswimmer, whose deformation is prescribed, 
can be generalized for a stochastic three-sphere microswimmer in the framework of non-equilibrium 
statistical mechanics. 
Using the concept of microrheology~\cite{MW95,GSOMS,FurstBook}, on the other hand, one can 
also discuss the locomotion of a three-sphere microswimmer in viscoelastic fluids or structured 
fluids~\cite{Yasuda17b,Yasuda20,Yasuda18}.
In this review article, we focus on the three-sphere microswimmer and give an overview of 
its various extensions that have been developed by the present authors and their collaborators. 
Notice that a comprehensive review of three-sphere microswimmers is not intended in this paper.

In Sect.~\ref{sec:original}, we first describe the original three-sphere swimmer proposed by  
Najafi and Golestanian~\cite{Golestanian04,Golestanian08}. 
In Sect.~\ref{sec:elastic}, we discuss an elastic three-sphere swimmer, in which the spheres 
are connected by two springs with time-dependent natural lengths~\cite{Yasuda17a}.
In Sect.~\ref{sec:thermal}, we consider the locomotion of a stochastic microswimmer, in which 
the three spheres have different temperatures~\cite{Hosaka17}. 
In Sect.~\ref{sec:odd}, we explain a three-sphere microswimmer, in which the spheres are 
connected by springs that exhibit odd elasticity~\cite{Yasuda21,Kobayashi23}.
In Sect.~\ref{sec:autonomous}, we propose an autonomous three-sphere microswimmer that 
can determine the velocity by itself in the steady state~\cite{Era21}.
In Sect.~\ref{sec:two}, we mention the hydrodynamic interaction between two elastic 
three-sphere microswimmers~\cite{Kuroda19}.
In Sect.~\ref{sec:viscoelastic}, we propose a new swimming mechanism for a three-sphere 
microswimmer in a viscoelastic fluid~\cite{Yasuda17b,Yasuda20}.
In Sect.~\ref{sec:structured}, we further investigate the effects of the intermediate structures of the 
surrounding viscoelastic fluid on the locomotion of a three-sphere microswimmer~\cite{Yasuda18}. 
In the final section, we shall provide a brief outlook on the other possible extensions of three-sphere 
microswimmers.

\section{Najafi-Golestanian Three-Sphere Microswimmer}
\label{sec:original}

\begin{figure*}[tbh]
\begin{center}
\includegraphics[scale=0.52]{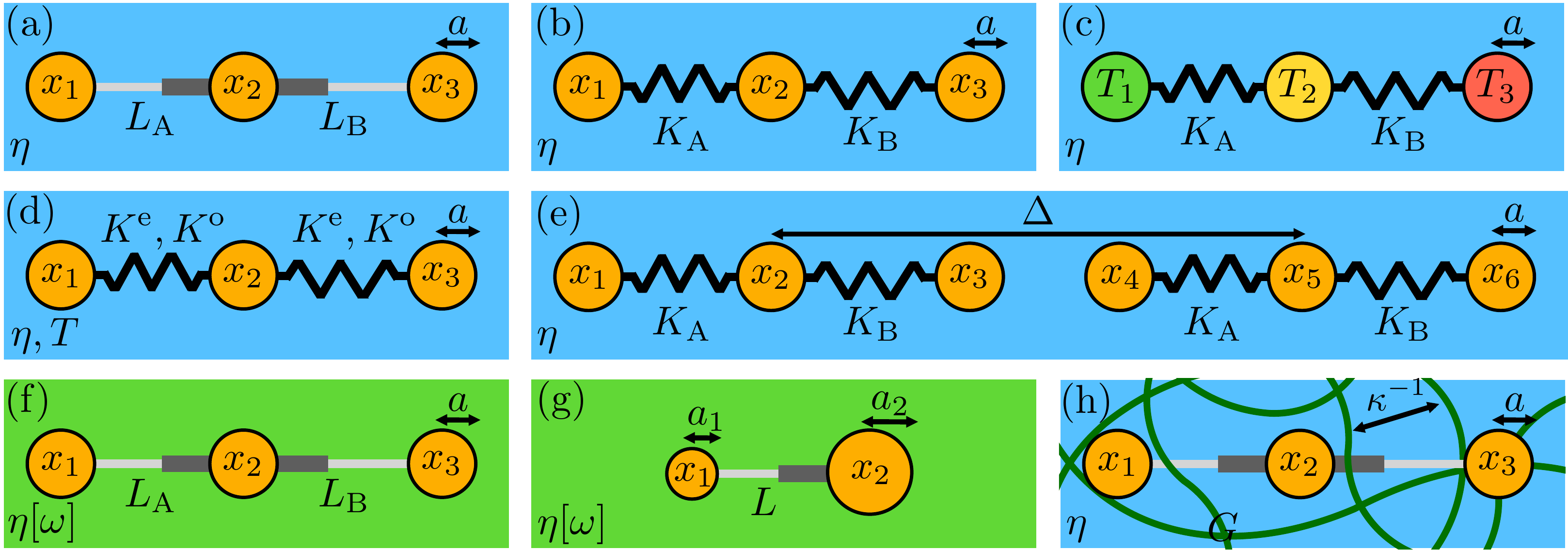}
\end{center}
\caption{
(Color online) 
(a)
Najafi-Golestanian three-sphere microswimmer model (see Sect.~\ref{sec:original}).
Three identical spheres of radius $a$ are connected by two arms of lengths $L_{\rm A}$ and 
$L_{\rm B}$ that undergo prescribed cyclic motions.
The positions of the spheres are denoted by $x_1$, $x_2$, and $x_3$ in a 
one-dimensional coordinate system. 
Such a microswimmer is embedded in a viscous fluid characterized by a constant shear viscosity $\eta$.
Throughout the paper, we use Roman subscripts $i, j = 1, 2, 3$ for the spheres and Greek subscripts 
$\alpha, \beta = \mathrm{A}, \mathrm{B}$ for the arms or the springs.
(b) 
Elastic three-sphere microswimmer in a viscous fluid (see Sect.~\ref{sec:elastic}). 
Three spheres are connected by two harmonic springs with elastic constants $K_{\rm A}$ 
and $K_{\rm B}$.  
The natural lengths of the springs, denoted by $\ell_{\rm A}$ and $\ell_{\rm B}$, undergo prescribed 
cyclic changes.
(c)
Thermal three-sphere microswimmer in a viscous fluid (see Sect.~\ref{sec:thermal}).
Three spheres are connected by two harmonic springs as in (b).
In this model, the three spheres are in equilibrium with independent heat baths having different 
temperatures $T_1$, $T_2$, and $T_3$.
Heat transfer between different spheres causes the locomotion of the microswimmer.
(d)
Odd three-sphere microswimmer in a viscous fluid having temperature $T$ (see Sect.~\ref{sec:odd}).
Three spheres are connected by two springs with both even elastic constant 
$K^\mathrm e$ and odd elastic constant $K^\mathrm o$. 
(The elastic constant $K_{\rm \alpha}$ in (b) and (c) corresponds to even elasticity.)
In this model, odd elasticity causes the non-reciprocal interaction between the two springs.
(e)
Two interacting elastic three-sphere microswimmers in a viscous fluid (see Sect.~\ref{sec:two}).
The positions of the three spheres in the left (L) swimmer are denoted by $x_1$, $x_2$, and $x_3$,
while those in the right (R) swimmer are denoted by $x_4$, $x_5$, and $x_6$.
The distance between the two swimmers is defined by $\Delta=x_5-x_2$. 
(f)
Najafi-Golestanian three-sphere microswimmer in a viscoelastic fluid characterized by a 
frequency-dependent complex shear viscosity $\eta[\omega]$ (see Sect.~\ref{sec:viscoelastic}).
The average velocity has both viscous and elastic contributions indicating the generalization of 
the scallop theorem for viscoelastic fluids.  
(g)
Asymmetric two-sphere microswimmer in a viscoelastic fluid (see Sect.~\ref{sec:viscoelastic}).
Two spheres having different radii $a_1$ and $a_2$ ($a_1<a_2$) are connected by an arm of 
length $L$.
Since there is only one degree of freedom, any periodic arm motion is reciprocal rather than 
non-reciprocal.
The average velocity of such a two-sphere microswimmer is nonzero. 
(h)
Najafi-Golestanian three-sphere microswimmer in a structured fluid that has intermediate 
mesoscopic structures (see Sect.~\ref{sec:structured}).
Within the two-fluid model, a polymer gel consists of an elastic network characterized by a  
shear modulus $G$ and a viscous fluid characterized by a viscosity $\eta$. 
Then the viscoelastic time scale is given by $\tau_{\rm v}=\eta/G$.
On the other hand, the elastic and fluid components are coupled to each other through mutual friction. 
Such friction introduces a characteristic length scale $\kappa^{-1}$ corresponding to the 
polymer mesh size.
The swimmer sizes ($a$ and $\ell$) compete with the characteristic length scale $\kappa^{-1}$. 
}
\label{fig1}
\end{figure*}

In this section, we shall first review the three-sphere microswimmer that was originally 
proposed by Najafi and Golestanian~\cite{Golestanian04} and later discussed in more detail by 
Golestanian and Ajdari~\cite{Golestanian08}. 
Consider a three-sphere microswimmer, in which the positions of the three spheres  
are given by $x_i$  ($i=1, 2, 3$) in a one-dimensional coordinate system as shown in Fig.~\ref{fig1}(a).
One can assume $x_1<x_2<x_3$ without loss of generality.
Although the size of the three spheres can be different in general~\cite{GolestanianEPJE2008,Nasouri19},
we mainly discuss the equal-size case of radius $a$ as it is sufficient to describe the essential 
swimming mechanism. 
The three spheres are connected by two arms of lengths $L_\alpha$ ($\alpha=\rm{A}, \rm{B}$) 
that can vary in time.
Along the swimmer axis, each sphere exerts a force $f_i$ on (and experiences a force $-f_i$ from) the 
fluid having a shear viscosity $\eta$. 
Because we are interested in autonomous net swimming, $f_i$ should satisfy the force-free condition,
i.e., $f_1+f_2+f_3=0$.

For $a/L_\alpha \ll 1$, the equations of motion for each sphere can be written as 
\begin{equation}
\dot x_i  = M_{ij}f_j,
\label{motion}
\end{equation}
where the dot indicates the time derivative, $\dot x_i=dx_i/dt$, and $M_{ij}$ is the 
mobility coefficient matrix describing the hydrodynamic interactions.
The summation over repeated indices ($i, j$ and $\alpha, \beta$) is assumed throughout this paper.
When the spheres are considerably far from each other ($a \ll \vert x_i-x_j \vert$),
$M_{ij}$ can be approximated as~\cite{Happel} 
\begin{equation}
M_{ij}=
\begin{cases}
1/(6\pi\eta a) & i=j, \\
1/(4\pi\eta \vert x_i-x_j \vert) & i\neq j.
\end{cases}
\label{condition}
\end{equation}
In the above, the case of $i=j$ corresponds to the Stokes mobility, whereas the case of $i \neq j$ 
describes the hydrodynamic interaction due to the Oseen tensor.
However, the tensorial structure of the Oseen tensor does not play a role in the present 
one-dimensional setup.
The total instantaneous velocity of the microswimmer is simply given by
$V=(\dot{x}_1+\dot{x}_2+\dot{x}_3)/3$.

One way to close the above equations is to prescribe the motion of the two arms.
In other words, the arm lengths $L_\alpha$ are known functions of time and they should satisfy 
the following relations:
\begin{align}
L_{\rm A}(t) = x_2(t)-x_1(t), \quad
L_{\rm B}(t) = x_3(t)-x_2(t).
\label{prescribedmotion}
\end{align} 
Then the total velocity can be obtained in terms of the arm lengths as~\cite{Golestanian08} 
\begin{align}
V=\frac{a}{6}\left[ \frac{\dot{L}_{\rm B}-\dot{L}_{\rm A}}{L_{\rm A}+L_{\rm B}}
+2 \left( \frac{\dot{L}_{\rm A}}{L_{\rm B}} -\frac{\dot{L}_{\rm B}}{L_{\rm A}} \right) \right].
\label{velocitygeneralnon}
\end{align}
For relatively small deformations, we can define the small displacements of the arms with 
respect to the average arm length $\ell$ as 
\begin{align}
u_{\rm A}(t) = x_2(t)-x_1(t) - \ell, \quad
u_{\rm B}(t) = x_3(t)-x_2(t) - \ell, 
\label{armexpnasion}
\end{align}  
where the condition $u_\alpha/\ell \ll 1$ is assumed.
These small displacements are related to the sphere velocities as 
$\dot u_{\rm A} = \dot x_2- \dot x_1$ and $\dot u_{\rm B} = \dot x_3- \dot x_2$.
Then, up to the leading order in $u_\alpha/\ell$, the average swimming velocity can be generally 
written as~\cite{Golestanian08}  
\begin{align}
\overline{V}=\frac{7a}{24\ell^2} 
\overline{(u_{\rm A} \dot{u}_{\rm B} - \dot{u}_{\rm A} u_{\rm B})},
\label{velocitygeneral}
\end{align}
where the averaging, indicated by the bar, is performed by time integration in a full cycle.
The above expression indicates that the average velocity is determined by the area enclosed by 
the orbit of periodic motion in the configuration space~\cite{Shapere89,Yasuda21catalytic}.
Hence, Eq.~(\ref{velocitygeneral}) can be understood as the mathematical expression of the
scallop theorem as long as the prescribed cyclic deformation is deterministic~\cite{Purcell77,Ishimoto12}.

As an example, let us consider the following harmonic deformations of the two arms 
\begin{align}
u_{\rm A}(t) = d_{\rm A} \cos (\Omega t), \quad
u _{\rm B}(t) = d_{\rm B} \cos (\Omega t -\phi),
\label{harmonicdeformation}
\end{align}  
where $d_\alpha$ is the amplitude, $\Omega$ is the common frequency, and $\phi$ is the 
phase difference between the two arms.  
The average swimming velocity in Eq.~(\ref{velocitygeneral}) now reads 
\begin{align}
\overline{V}=\frac{7a d_{\rm A} d_{\rm B} \Omega}{24\ell^2} \sin \phi,
\label{velocityharmonic}
\end{align}
which is proportional to $\Omega$.
This result clearly shows that $\overline{V}$ is nonzero when $\phi \neq 0, \pi$ for which the arm motion 
is non-reciprocal.
The swimming direction is determined by the sign of $\sin \phi$, and the maximum velocity is obtained 
when the phase difference is $\phi =\pm \pi/2$.
We also remind that $\overline{V}$ in Eq.~(\ref{velocityharmonic}) does not depend on the viscosity 
$\eta$ because we have prescribed the motion of the two arms.
This situation will be modified in the models discussed later.

\section{Elastic Three-Sphere Microswimmer}
\label{sec:elastic}

In this section, we discuss the first generalization of a three-sphere microswimmer, in which the spheres 
are connected by two harmonic springs, i.e., an elastic three-sphere microswimmer~\cite{Yasuda17a}.
In this model, the natural length of each spring depends on time and is assumed to undergo a prescribed 
cyclic change. 
Introducing harmonic springs between the spheres leads to an intrinsic time scale of an elastic microswimmer
characterizing its internal relaxation dynamics.

As schematically shown in Fig.~\ref{fig1}(b), the present model consists of three hard spheres connected 
by two harmonic springs A and B with spring constants 
$K_{\rm A}$ and $K_{\rm B}$, respectively.
We assume that the natural lengths of these springs, denoted by $\ell_{\rm A}(t)$ and 
$\ell_{\rm B}(t)$, undergo cyclic time-dependent change.
Since the energy of an elastic microswimmer is given by 
\begin{align}
E = \frac{K_{\rm A}}{2}(x_2 - x_1 - \ell_{\rm A})^2 + \frac{K_{\rm B}}{2}(x_3 - x_2 - \ell_{\rm B})^2,
\end{align}
the three forces $f_i=-\partial E/\partial x_i$ in Eq.~(\ref{motion}) read 
\begin{align}
f_1&=K_{\rm A}(x_2-x_1-\ell_{\rm A}),
\label{force1} \\
f_2&=-K_{\rm A}(x_2-x_1-\ell_{\rm A})+K_{\rm B}(x_3-x_2-\ell_{\rm B}),
\label{force2} \\
f_3&=-K_{\rm B}(x_3-x_2-\ell_{\rm B}).
\label{force3} 
\end{align}
Notice that the force-free condition, $f_1+f_2+f_3=0$, is automatically satisfied in this model.

Next, we assume that the two natural lengths of the springs undergo the following periodic changes: 
\begin{align}
\ell_{\rm A}(t) =\ell+d_{\rm A}\cos(\Omega t), \quad 
\ell_{\rm B}(t) =\ell+d_{\rm B}\cos(\Omega t - \phi).
\label{ellA}
\end{align}
Here, $\ell$ is the constant natural length, $d_{\alpha}$ is the amplitude of the oscillatory change, 
$\Omega$ is the common frequency, and $\phi$ is the phase difference between 
the two cyclic changes.
It is convenient to introduce the characteristic time scale $\tau=6\pi\eta a/K_{\rm A}$ and define 
the scaled frequency by $\hat \Omega = \Omega \tau$. 
We also denote the ratio between the two spring constants by $\lambda = K_{\rm B}/K_{\rm A}$.

\begin{figure}[t]
\begin{center}
\includegraphics[scale=0.5]{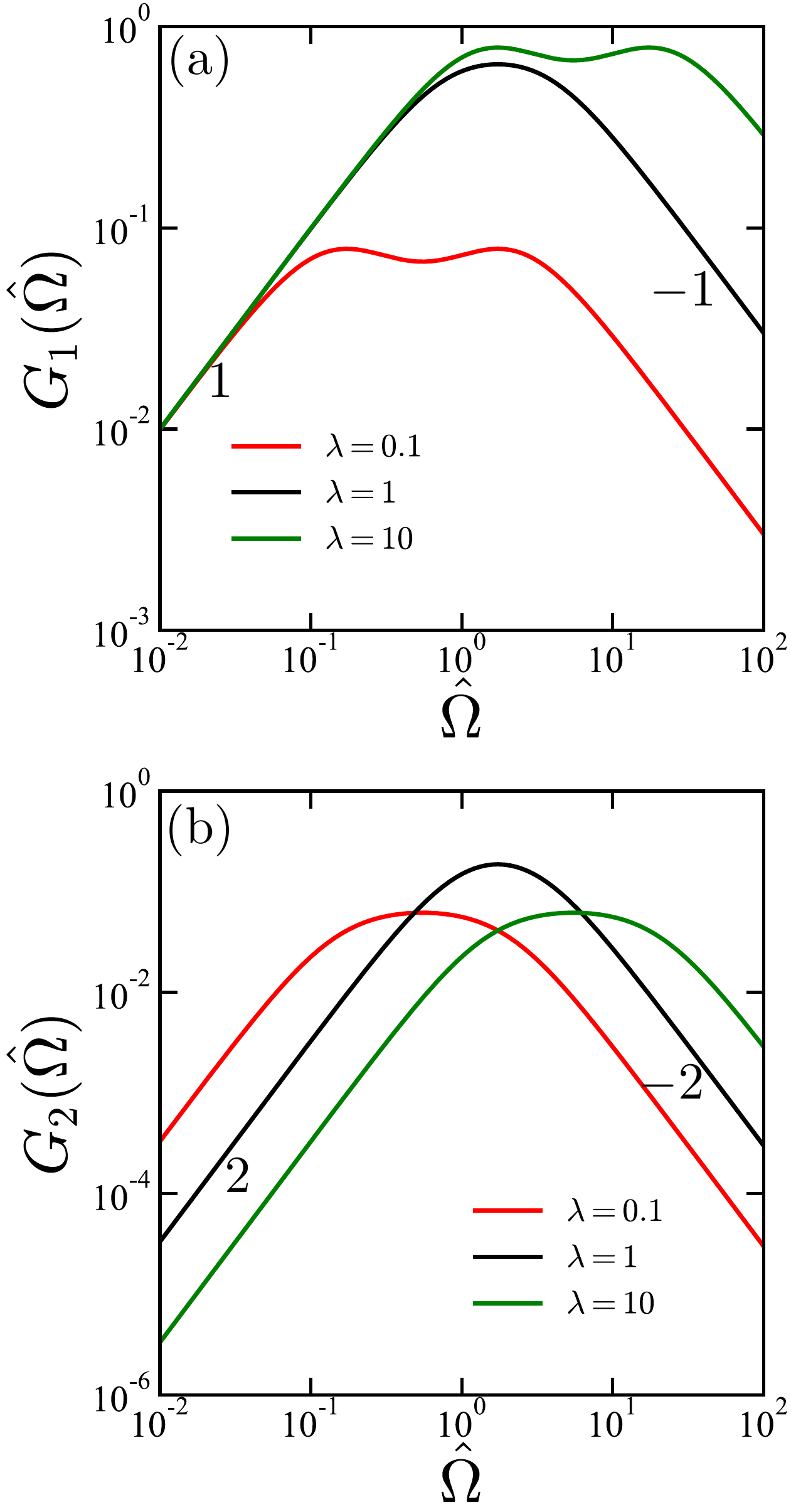}
\end{center}
\caption{
(Color online) Plots of the scaling functions (a) $G_1(\hat \Omega; \lambda)$ and (b) 
$G_2(\hat \Omega; \lambda)$ defined in Eqs.~(\ref{F1}) and (\ref{F2}), respectively, 
as functions of $\hat \Omega = \Omega \tau$ ($\tau=6\pi\eta a/K_{\rm A}$)
for $\lambda= K_{\rm B}/K_{\rm A}=0.1, 1,$ and $10$.
The numbers indicate the slope representing the exponent of the power-law behaviors.
Reprinted from Ref.~\cite{Yasuda17a}.
$\copyright$ 2017 The Physical Society of Japan.}
\label{fig2}
\end{figure}

Similar to Eq.~(\ref{armexpnasion}), we define the spring extensions $u_\alpha$ relative to $\ell$.
Then, by using Eq.~(\ref{velocitygeneral}), we obtain the average velocity up to the lowest-order terms as 
\begin{align}
\overline V & =\frac{7 a d_{\rm A}d_{\rm B}}{24\ell^2\tau}
G_1(\hat \Omega;\lambda)\sin\phi
+\frac{7 (1-\lambda)a d_{\rm A} d_{\rm B}}{12\ell^2 \tau}
G_2(\hat \Omega;\lambda) \cos\phi
\nonumber \\
& +\frac{7a (d_{\rm A}^2-d_{\rm B}^2\lambda)}{24\ell^2\tau}
G_2(\hat \Omega;\lambda),
\label{generalvelocity}
\end{align}
where the two scaling functions are given by 
\begin{align}
G_1(\hat \Omega;\lambda) &=
\frac{3\lambda \hat \Omega(3\lambda+\hat \Omega^2)}
{9\lambda^2+2(2+\lambda+2\lambda^2)\hat \Omega^2+\hat \Omega^4},
\label{F1}
\\
G_2(\hat \Omega;\lambda) & 
=\frac{3\lambda \hat\Omega^2}{9\lambda^2+2(2+\lambda+2\lambda^2)\hat \Omega^2+\hat \Omega^4}.
\label{F2}
\end{align}
In Fig.~\ref{fig2}, we plot the above scaling functions as functions of $\hat \Omega$ for 
$\lambda=0.1, 1$ and $10$. 
Notice, however, that the cases $\lambda=0.1$ and $10$ are essentially equivalent because we 
can always exchange the springs A and B, whereas we have defined the relaxation time $\tau$ 
by using $K_{\rm A}$ (and not $K_{\rm B}$).

For the symmetric case when $\lambda=1$ and $d_{\rm A}=d_{\rm B}=d$, only the first term in 
Eq.~(\ref{generalvelocity}) remains. 
In this case, we have 
\begin{align}
\overline V= \frac{7a d^2}{24\ell^2 \tau}    
\frac{3\hat \Omega(3+ \hat \Omega^2)}{9+10 \hat \Omega^2+ \hat \Omega^4} \sin\phi.
\label{symvelocity}
\end{align}
In the small-frequency limit of $\hat \Omega \ll 1$, the average velocity increases as 
$\overline V \sim \Omega$ and coincides with Eq.~(\ref{velocityharmonic}).
This is because small $\hat \Omega$ corresponds to large $K_{\rm A}$ and the springs behave as 
rigid arms. 
In the opposite large-frequency limit of $\hat \Omega \gg 1$, on the other hand, the average velocity 
decreases as $\overline V \sim \Omega^{-1}$ as $\Omega$ is increased.
When $K_{\rm A}$ is small, it takes time for the spring to relax to its natural length, which delays the 
mechanical response. 
The crossover frequency between the above two regimes is given by $\hat \Omega^{\ast} \approx 1$.

When $\lambda \neq 1$, on the other hand, the second term in Eq.~(\ref{generalvelocity}) is present even 
if $\phi=0$.
The third term is also present when $d_{\rm A}^2 \neq d_{\rm B}^2\lambda$, regardless 
of the $\phi$-value.
In contrast to the first term representing the non-reciprocal arm motion, both the second and third 
terms reflect the structural asymmetry of an elastic three-sphere microswimmer. 
The frequency dependence of the second and third terms, represented by 
$G_2(\hat \Omega, \lambda)$, differs from that of the first term, represented by 
$G_1(\hat \Omega, \lambda)$.
From Eq.~(\ref{F2}), we see that $\overline V$ due to the second and third terms increases 
as $\overline{V} \sim \Omega^2$ for $\hat \Omega \ll 1$, whereas it decreases as 
$\overline{V} \sim \Omega^{-2}$ for $\hat \Omega \gg 1$.

In general, the overall swimming velocity depends on various structural parameters and 
exhibits a complex frequency dependence. 
For example, $G_1(\hat \Omega, \lambda)$ in Fig.~\ref{fig2}(a) 
exhibits a non-monotonic frequency dependence (two maxima) for $\lambda=0.1$ and $10$.
On the other hand, an important common feature for all the terms in Eq.~(\ref{generalvelocity}) 
is that $\overline V$ decreases for large $\hat \Omega$, which is a characteristic feature 
of an elastic three-sphere microswimmer.

\section{Thermal Three-Sphere Microswimmer}
\label{sec:thermal}

Extending the model of an elastic three-sphere microswimmer, we propose a different locomotion 
mechanism that is purely induced by thermal fluctuations~\cite{Hosaka17}. 
Here, the key assumption is that the three spheres are in equilibrium with independent heat baths 
at different temperatures.
Then, the heat transfer occurs from a hotter sphere to a colder one, driving the whole system 
out of equilibrium. 
Since this model is similar to a class of thermal ratchet models, the suggested mechanism can be 
relevant to the non-equilibrium dynamics of proteins and enzymes in biological systems~\cite{Hosaka22}.

As shown in Fig.~\ref{fig1}(c), we consider an elastic three-sphere microswimmer, in which 
the three spheres are in equilibrium with independent heat baths having temperatures $T_i$.
When these temperatures are different, the system is driven out of equilibrium
because a heat flux is generated from a hotter sphere to a colder one.
In the presence of thermal fluctuations, the equations of motion of the three spheres can be 
written as~\cite{DoiBook}  
\begin{equation}
\dot x_i  = M_{ij}f_j + \xi_i,
\label{motionnoise}
\end{equation}
where the mobility matrix $M_{ij}$ is given by Eq.~(\ref{condition}) and the forces $f_i$ are 
given by Eqs.~(\ref{force1})-(\ref{force3}) as before.
The additional terms describing the white-noise sources $\xi_i(t)$ have zero mean, i.e., 
$\langle \xi_i(t) \rangle =0$, and their correlations satisfy
\begin{align}
\langle \xi_i(t)\xi_j(t')\rangle =2 D_{ij} \delta(t-t'), 
\label{FDTreal}
\end{align}
where $\langle \cdots \rangle$ indicates the ensemble average, namely, the average for 
many equivalent systems, 
In the above, $D_{ij}$ is the mutual diffusion coefficient given by
\begin{equation}
D_{ij}=
\begin{cases}
k_{\rm B}T_i/(6 \pi \eta a) & i=j, \\
k_{\rm B}\Theta(T_i, T_j)/(4 \pi \eta \vert x_i -x_j \vert) & i\neq j,
\end{cases}
\label{diffusionmatrix}
\end{equation}
where $k_{\rm B}$ is the Boltzmann constant and $\Theta(T_i, T_j)$ is a function of $T_i$ and $T_j$. 
The relevant effective temperature can be the mobility-weighted average~\cite{Grosberg15}, which 
in the present case is simply given by $\Theta(T_i, T_j)=(T_i + T_j)/2$
because all the spheres have the same size.
However, its explicit functional form is not needed here, and we only require that 
$\Theta$ should satisfy an appropriate fluctuation-dissipation theorem in thermal equilibrium.
This is allowed because we only consider the limit of $a \ll \ell$.

The above stochastic equations can be solved in the Fourier domain. 
Assuming $u_\alpha \ll \ell$ and $a \ll \ell$, we obtain the lowest-order average velocity as~\cite{Hosaka17} 
\begin{align}
\langle V\rangle & =\frac{k_{\rm B}}{144\pi\eta\ell^2(1+\lambda)}
[(2 -5 \lambda)T_1 
\nonumber \\
& -7(1 - \lambda)T_2+(5-2\lambda)T_3],
\label{Vgeneral}
\end{align}
where $\lambda = K_{\rm B}/K_{\rm A}$ as before.
When the three temperatures are identical ($T_1=T_2=T_3$), the velocity vanishes
identically, $\langle V\rangle =0$.
This indicates that a thermal three-sphere microswimmer can acquire a finite velocity 
owing to the temperature difference among the spheres.

When the springs are symmetric and $\lambda=1$, Eq.~(\ref{Vgeneral}) reduces to 
\begin{align}
\langle V\rangle=\frac{k_{\rm B}(T_3-T_1)}{96\pi\eta\ell^2},
\label{Vsymmetric}
\end{align}
which is proportional to the temperature difference $T_3-T_1$.
Since we have assumed $x_1<x_2<x_3$, the swimming direction is from a colder 
sphere to a hotter one ($\langle V \rangle >0$) when $T_3>T_1$ 
and vice versa. 
It is also remarkable that Eq.~(\ref{Vsymmetric}) does not depend on the temperature 
$T_2$ of the middle sphere.  
Hence $\langle V \rangle =0$ when $T_1=T_3$ even though $T_1$ and $T_3$ are 
different from $T_2$. 
However, the presence of the middle sphere is essential for directional locomotion because 
the hydrodynamic interactions among the three spheres are responsible for the motion.

The analytically obtained velocity in Eq.~(\ref{Vgeneral}) can be related to the heat flows in the 
steady state. 
Following Ref.~\cite{SekimotoBook} and retaining up to the lowest-order terms, we obtain the 
average heat gain per unit time for each sphere as 
\begin{align}
\langle \dot{Q}_1 \rangle & =  
\frac{k_{\rm B}}{6\tau(1+\lambda)}
[(3 + 2 \lambda)T_1 -(3+\lambda)T_2 -\lambda T_3],
\label{Q1}
\\
\langle \dot{Q}_2 \rangle & =  
\frac{k_{\rm B}}{6\tau(1+\lambda)}
[-(3 +\lambda) T_1+(3 +2\lambda+3\lambda^2)T_2 
\nonumber \\
&-(\lambda+3\lambda^2)T_3],
\label{Q2}
\\
\langle \dot{Q}_3 \rangle & =  
\frac{k_{\rm B}}{6\tau(1+\lambda)}
[-\lambda T_1 -(\lambda+3 \lambda^2)T_2
+(2 \lambda+3 \lambda^2)T_3],
\label{Q3}
\end{align}
which all vanish when $T_1=T_2=T_3$.
Notice that the above heat flows also satisfy
$\langle \dot{Q}_1 \rangle+\langle \dot{Q}_2 \rangle
+\langle \dot{Q}_3 \rangle=0$.
Assuming a linear relationship between the average velocity in Eq.~(\ref{Vgeneral}) and 
the heat flows in Eqs.~(\ref{Q1})-(\ref{Q3}), we obtain an alternative expression 
for the velocity:
\begin{align}
\langle V \rangle =\frac{a}{8K_{\rm A}\ell^2}
\left[ 
\frac{3-5\lambda}{1+\lambda} \langle \dot{Q}_1 \rangle
+ \frac{5-3\lambda}{\lambda(1+\lambda)}\langle \dot{Q}_3 \rangle
\right].
\label{vel-heat}
\end{align}
For the symmetric case of $\lambda=1$ corresponding to Eq.~(\ref{Vsymmetric}), 
the above expression reduces to
\begin{align}
\langle V \rangle=\frac{a}{8K_{\rm A}\ell^2}\left[\langle \dot{Q}_3 \rangle-
\langle \dot{Q}_1\rangle \right]. 
\label{vel-heat-sym}
\end{align}
This relation indicates that the net heat flow between the first and third spheres determines 
the average velocity.

Previously, Yang \textit{et al.}\ performed hydrodynamic simulations of a self-thermophoretic Janus 
particle~\cite{Yang14} and they reproduced the experimental observation by Jiang \textit{et al.}~\cite{Jiang10}.
The above thermal three-sphere microswimmer is different because thermal fluctuations of internal 
degrees of freedom cause locomotion.
We also note from Eq.~(\ref{Vgeneral}) that $\langle V \rangle$ is nonzero for symmetric 
temperatures $T_1=T_3 \neq T_2$ as long as the structural asymmetry exists ($\lambda \neq 0$), 
which cannot be realized for a thermophoretic Janus particle.

\section{Odd Three-Sphere Microswimmer}
\label{sec:odd}

Recently, Scheibner \textit{et al.}\ introduced the concept of odd elasticity that 
arises from non-reciprocal interactions in active systems~\cite{Scheibner20,Fruchart23}.
The odd part of the elastic constant matrix quantifies the amount of 
work extracted along quasistatic deformation cycles.
In this section, we discuss another type of thermally driven microswimmer, in which 
the three spheres are connected by odd springs~\cite{Yasuda21,Kobayashi23}.

As shown in Fig.~\ref{fig1}(d),
consider an elastic three-sphere microswimmer, in which the three spheres are connected 
by two identical springs that exhibit both even and odd elasticity.
Then, the forces $F_\mathrm A$ and $F_\mathrm B$ conjugate to the spring extensions 
$u_\mathrm A$ and $u_\mathrm B$ [see Eq.~(\ref{armexpnasion})], respectively, are given by 
$F_\alpha=-K_{\alpha\beta}u_\beta$.
For an odd spring, the elastic constant matrix $K_{\alpha\beta}$ is given 
by~\cite{YasudaOM22,YasudaTCF22,KobayashiODD23,Ishimoto22}
\begin{align}
K_{\alpha\beta}=K^\mathrm{e}\delta_{\alpha\beta}+K^\mathrm{o}\epsilon_{\alpha\beta}.
\label{ElasticConstant}
\end{align}
Here, $K^\mathrm{e}$ and $K^\mathrm{o}$ are the even and odd elastic constants, respectively,
in the two-dimensional configuration space spanned by $u_\mathrm{A}$ and $u_\mathrm{B}$,
$\delta_{\alpha\beta}$ is the Kronecker delta, and $\epsilon_{\alpha\beta}$ is the 
two-dimensional Levi-Civita tensor with $\epsilon_{\mathrm{AA}}=\epsilon_{\mathrm{BB}}=0$ and 
$\epsilon_{\mathrm{AB}}=-\epsilon_{\mathrm{BA}}=1$.
(The spring constants $K_{\rm A}$ and $K_{\rm B}$ in Sect.~\ref{sec:elastic} correspond to 
even elastic constants.)
The presence of the odd elasticity $K^\mathrm{o}$ in Eq.~(\ref{ElasticConstant}) 
reflects the non-reciprocal interaction between the two springs such that $u_\mathrm A$ and $u_\mathrm B$ 
influence each other in a different way.
The forces $f_i$ acting on each sphere are given by 
$f_1=-F_\mathrm A$, $f_2=F_\mathrm A-F_\mathrm B$, and $f_3=F_\mathrm B$.

Such an odd microswimmer is immersed in a fluid with a shear viscosity of $\eta$ and temperature $T$. 
Similar to the previous stochastic model, the equations of motion for each sphere are given by 
Eq.~(\ref{motionnoise}).
In the current model, the Gaussian white-noise sources $\xi_i$ also have zero mean 
$\langle\xi_i (t) \rangle=0$, and their correlations satisfy the following 
fluctuation-dissipation theorem:
\begin{align}
\langle\xi_i(t)\xi_j(t')\rangle=2k_\mathrm B T M_{ij}\delta(t-t'),
\label{FDT}
\end{align}
where $M_{ij}$ is given by Eq.~(\ref{condition}).

\begin{figure}[t]
\begin{center}
\includegraphics[scale=0.38]{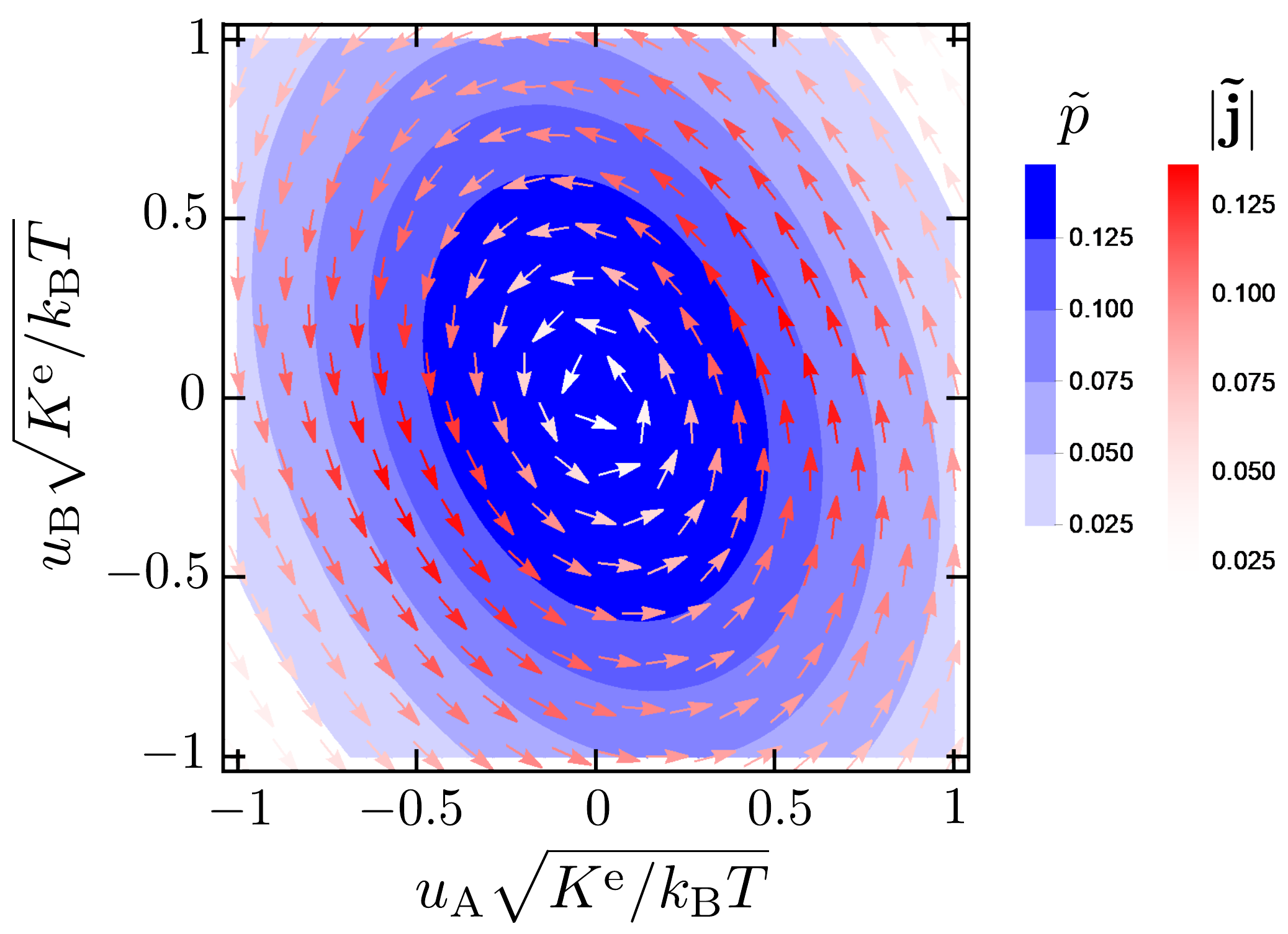}
\end{center}
\caption{(Color online) Steady-state scaled probability distribution function 
$\tilde p=p k_\mathrm{B}T/K^\mathrm{e}$ [see Eq.~(\ref{PDF})] 
and scaled probability flux 
$\tilde{\mathbf{j}}=\mathbf{j} \tau\sqrt{k_\mathrm{B}T/K^\mathrm{e}}$ 
[arrows, see Eq.~(\ref{Flux})] ($\tau=6\pi\eta a/K^\mathrm e$) in the configuration space spanned by 
$u_\mathrm{A}$ and $u_\mathrm{B}$ when $\nu=K^\mathrm o/K^\mathrm e=1$.
Reproduced from Ref.~\cite{Yasuda21}.
$\copyright$ 2021 K. Yasuda, Y. Hosaka, I. Sou, and S. Komura.
}
\label{fig3}
\end{figure}

The equal-time correlation functions can be obtained from 
the reduced Langevin equations for $\dot u_\mathrm A=\dot x_2-\dot x_1$ and 
$\dot u_\mathrm B=\dot x_3-\dot x_2$ as 
\begin{align}
\dot u_\alpha=\Gamma_{\alpha\beta}u_\beta+\Xi_\alpha.
\label{DynamicEqu}
\end{align}
In the above, $\Gamma_{\alpha\beta}$ and $\Xi_\alpha$ are 
\begin{align}
\mathbf{\Gamma} =  -\frac{1}{\tau} \left(
    \begin{array}{ccc}
      2+\nu & -1+2\nu \\
      -1-2\nu & 2-\nu
    \end{array}
  \right), \quad 
\mathbf{\Xi} =  \left(
    \begin{array}{cc}
      \xi_2-\xi_1 \\
      \xi_3-\xi_2
    \end{array}
  \right),
  \label{GammaXi}
\end{align}
where $\tau=6\pi\eta a/K^\mathrm e$ and we have introduced the ratio $\nu=K^\mathrm o/K^\mathrm e$.
Notice that $\Gamma_{\alpha\beta}$ is non-reciprocal, i.e., $\Gamma_\mathrm{AB}\neq \Gamma_\mathrm{BA}$
when $\nu \neq 0$.
By solving Eq.~(\ref{DynamicEqu}) in the Fourier domain and using Eq.~(\ref{FDT}), we can calculate the 
equal-time correlation functions $\langle u_\mathrm A^2 \rangle$, 
$\langle u_\mathrm B^2 \rangle$, and $\langle u_\mathrm Au_\mathrm B \rangle$. 
From these quantities, we obtain the average velocity as
\begin{align}
\langle V\rangle=\frac{7k_\mathrm BT\nu}{48\pi \eta \ell^2}.
\label{velocity}
\end{align}
We see here that $\langle V\rangle$ is proportional to the odd elastic constant $K^\mathrm{o}$ that can 
take both positive and negative values.

Next, we discuss the non-equilibrium statistical properties of the odd three-sphere 
microswimmer~\cite{Weiss03,Weiss07}.
Consider the time-dependent probability distribution function $p(u_\mathrm{A},u_\mathrm{B},t)$.
The Fokker-Planck equation that corresponds to Eq.~(\ref{DynamicEqu}) can be written as 
$\dot p=-\partial_\alpha j_\alpha$,
where $\partial_\alpha=\partial/(\partial u_\alpha)$ and $j_\alpha$ is the probability flux given by~\cite{Sou19,Sou21} 
\begin{align}
j_\alpha=\Gamma_{\alpha\beta}u_\beta p-\mathcal{D}_{\alpha\beta} \partial_\beta p.
\label{Flux}
\end{align}
Here, $\mathcal{D}_{\alpha\beta}$ is the diffusion matrix 
\begin{align}
\mathbfcal{D}=  \frac{k_\mathrm BT}{6\pi\eta a}\left(
    \begin{array}{cc}
      2& -1\\
      -1&2
    \end{array}
  \right),
  \label{DM}
\end{align}
that satisfies the fluctuation-dissipation relation $\langle\Xi_\alpha(t)\Xi_\beta(t')\rangle=2\mathcal{D}_{\alpha\beta}\delta(t-t')$ 
because of Eq.~(\ref{FDT}) [note that $\mathcal{D}_{\alpha\beta}$ is different from $D_{ij}$ in Eqs.~(\ref{FDTreal})].

Owing to the reproductive property of Gaussian distributions, the steady-state probability distribution 
function that satisfies $\dot p=0$ is given by the following Gaussian function
\begin{align}
p(u_\mathrm{A},u_\mathrm{B})=\frac{1}{2\pi \sqrt{\det \mathbf C}}
\exp\left[-\frac{1}{2}(C^{-1})_{\alpha\beta}u_\alpha u_\beta\right].
\label{PDF}
\end{align}
Here, $C_{\alpha\beta}=\langle u_\alpha u_\beta\rangle$ is the covariance matrix given by  
\begin{align}
\mathbf{C}=\frac{k_\mathrm BT}{K^\mathrm e}\frac{1}{1+\nu^2}  \left(
    \begin{array}{cc}
      1-\nu/2+\nu^2 & -\nu^2/2 \\
      -\nu^2/2 & 1+\nu/2+\nu^2
    \end{array}
  \right),
  \label{CVM}
\end{align}
and $(C^{-1})_{\alpha\beta}$ is the inverse matrix of $C_{\alpha\beta}$.
Then the determinant of $\mathbf{C}$ becomes 
\begin{align}
\det \mathbf{C}=\left(\frac{k_\mathrm BT}{K^\mathrm e}\right)^2\frac{4+7\nu^2+3\nu^4}{4(1+\nu^2)^2}.
\label{detCVM}
\end{align}

In Fig.~\ref{fig3}, we plot the steady-state probability distribution function [Eq.~(\ref{PDF})] and  
the corresponding probability flux [Eq.~(\ref{Flux})] when $\nu=1$.
The probability distribution function is distorted by the negative correlation 
($C_\mathrm{AB}=C_\mathrm{BA}\sim -\nu^2/2$) between $u_\mathrm{A}$ and $u_\mathrm{B}$.
Importantly, one can see a counter-clockwise loop of the probability flux.
Such a probability flux becomes clockwise for $\nu <0$ and vanishes when $\nu=0$. 
Generally, the existence of the probability flux loop indicates that the detailed balance is 
broken in the non-equilibrium steady state~\cite{Battle16,Gnesotto18}.
In contrast to the deterministic scallop theorem mentioned in Eq.~(\ref{velocitygeneral}),
the presence of the probability flux loop can be regarded as the stochastic scallop theorem 
that can be applied to thermally driven microswimmers~\cite{Hosaka17,Yasuda21,Kobayashi23,Sou19,Sou21,KobayashiODD23}.

The steady-state probability flux can be conveniently expressed in terms of a frequency 
matrix $\Omega_{\alpha\beta}$ as $j_\alpha=\Omega_{\alpha\beta} u_\beta p$~\cite{Weiss03,Weiss07}.
In the current model, the frequency matrix is given by  
\begin{align}
\mathbf{\Omega}=\frac{3\nu}{\tau(4+3\nu^2)}  \left(
\begin{array}{ccc}
-\nu^2 & -2+\nu-2\nu^2 \\
2+\nu+2\nu^2 & \nu^2
\end{array}
\right),
\end{align}
which is traceless.
Then, the two eigenvalues of $\Omega_{\alpha\beta}$ are given by 
\begin{align}
\gamma=\pm {\mathrm i} \frac{3\nu}{\tau(4+3\nu^2)} \sqrt{4+7\nu^2+3\nu^4}.
\label{eigenvalue}
\end{align}
Comparing Eq.~(\ref{velocity}) with Eqs.~(\ref{detCVM}) and (\ref{eigenvalue}), we obtain the following
alternative expression for the absolute value of the average velocity: 
\begin{align}
\vert \langle V\rangle \vert =\frac{7a}{12\ell^2}\sqrt{\det \mathbf C}\,\vert \gamma \vert.
\end{align}
Here, $7a/(12\ell^2)$ is the geometrical factor,
$\sqrt{\det \mathbf C}\sim k_\mathrm BT/K^\mathrm e$ is the randomly explored area in the configuration 
space, and $\vert \gamma \vert \sim \tau^{-1}$ is the frequency of the rotational probability flux.
The above expression clarifies the physical meaning of the average velocity of an odd three-sphere 
microswimmer driven by thermal fluctuations.

\section{Autonomous Three-Sphere Microswimmer}
\label{sec:autonomous}

In this section, we propose a model of a three-sphere swimmer that can autonomously determine 
its velocity~\cite{Era21}.
To implement such a control mechanism, we introduce a coupling between the two natural 
lengths of an elastic microswimmer by using the interaction adopted in the Kuramoto model for 
coupled oscillators~\cite{Kuramoto84,Ritort05}.
Such a microswimmer acquires a steady-state velocity and a finite phase difference in 
the long-time limit without any external control.

Consider a symmetric elastic three-sphere microswimmer ($\lambda=1$), 
in which the natural lengths of the springs undergo the following cyclic changes in time
\begin{align}
\ell_{\rm A}(t) =\ell+d \cos [ \theta_{\rm A}(t) ], \quad 
\ell_{\rm B}(t) =\ell+d \cos [ \theta_{\rm B}(t) ],
\label{ellauto}
\end{align}
where $\theta_{\rm A}(t)$ and $\theta_{\rm B}(t)$ are the time-dependent phases.
Although these motions are the generalization of Eq.~(\ref{ellA}), the important feature 
is that $\theta_{\rm A}(t)$ and $\theta_{\rm B}(t)$ are affected by the relative positions and the 
velocities of the three spheres.
We employ the following time-evolution equations for $\theta_{\rm A}(t)$ and 
$\theta_{\rm B}(t)$ that describe a synchronization behavior~\cite{Kuramoto84,Ritort05}
\begin{align}
\dot{\theta}_{\rm A}&=\Omega+\chi \sin[\theta_{\rm A}(t)-\phi_{\rm A}(t)], 
\label{time_evolution_phase1}\\
\dot{\theta}_{\rm B}&=\Omega+\chi \sin[\theta_{\rm B}(t)-\phi_{\rm B}(t)],
\label{time_evolution_phase2}
\end{align}
where $\Omega$ is the constant frequency, 
$\chi \ge 0$ is the coupling parameter representing the strength of synchronization, 
and $\phi_{\rm A}$ and $\phi_{\rm B}$ are the mechanical phases as explained below.

\begin{figure}[t]
\centering
\includegraphics[scale=0.23]{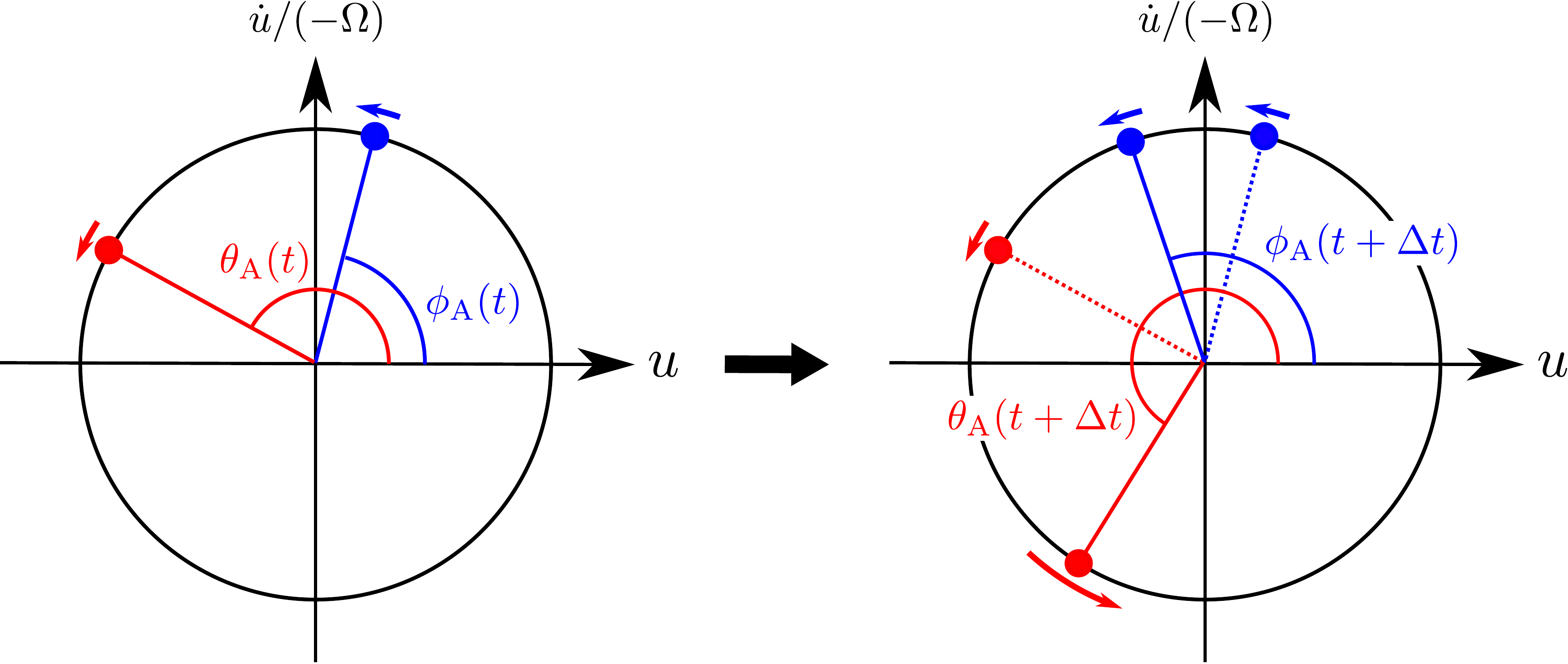}
\caption{
(Color online) Dynamics of $\theta_{\rm A}$ describing the phase of the natural length [see Eq.~(\ref{ellauto})]
and $\phi_{\rm A}$ describing the mechanical phase [see Eq.~(\ref{phiA})].
When $\theta_{\rm A} > \phi_{\rm A}$ at $t$, as shown in the left figure, and when $\chi>0$ in 
Eq.~(\ref{time_evolution_phase1}), the velocity $\dot{\theta}_{\rm A}$ becomes larger at a later time 
$t + \Delta t$, as shown in the right figure.
As a result, the difference between $\theta_{\rm A}$ and $\phi_{\rm A}$ also increases at $t + \Delta t$.
A similar dynamics occurs also for $\theta_{\rm B}$ and $\phi_{\rm B}$.
Reprinted from Ref.~\cite{Era21}.
$\copyright$ 2021 Institute of Physics.
}  
\label{fig4}
\end{figure}

To discuss the above mechanical phases, we use the 
spring displacements $u_{\rm A}$ and $u_{\rm B}$ defined in Eq.~(\ref{armexpnasion}).
Then the mechanical phases $\phi_{\rm A}$ and $\phi_{\rm B}$ are  
introduced by the relative positions and the velocities of the spheres as 
\begin{align}
\cos \phi_{\rm A}&=u_{\rm A}/U_{\rm A}, \quad 
\sin \phi_{\rm A}=-\dot{u}_{\rm A}/(\Omega U_{\rm A}), 
\label{phiA} \\
\cos \phi_{\rm B}&=u_{\rm B}/U_{\rm B}, \quad
\sin \phi_{\rm B}=-\dot{u}_{\rm B}/(\Omega U_{\rm B}), 
\label{phiB}
\end{align}
where $U_{\rm A(B)}=[u_{\rm A(B)}^{2}+(\dot{u}_{\rm A(B)}/\Omega)^2]^{1/2}$. 
The above equations complete the model for an autonomous three-sphere microswimmer.

\begin{figure}[tbh]
\centering
\includegraphics[scale=0.35]{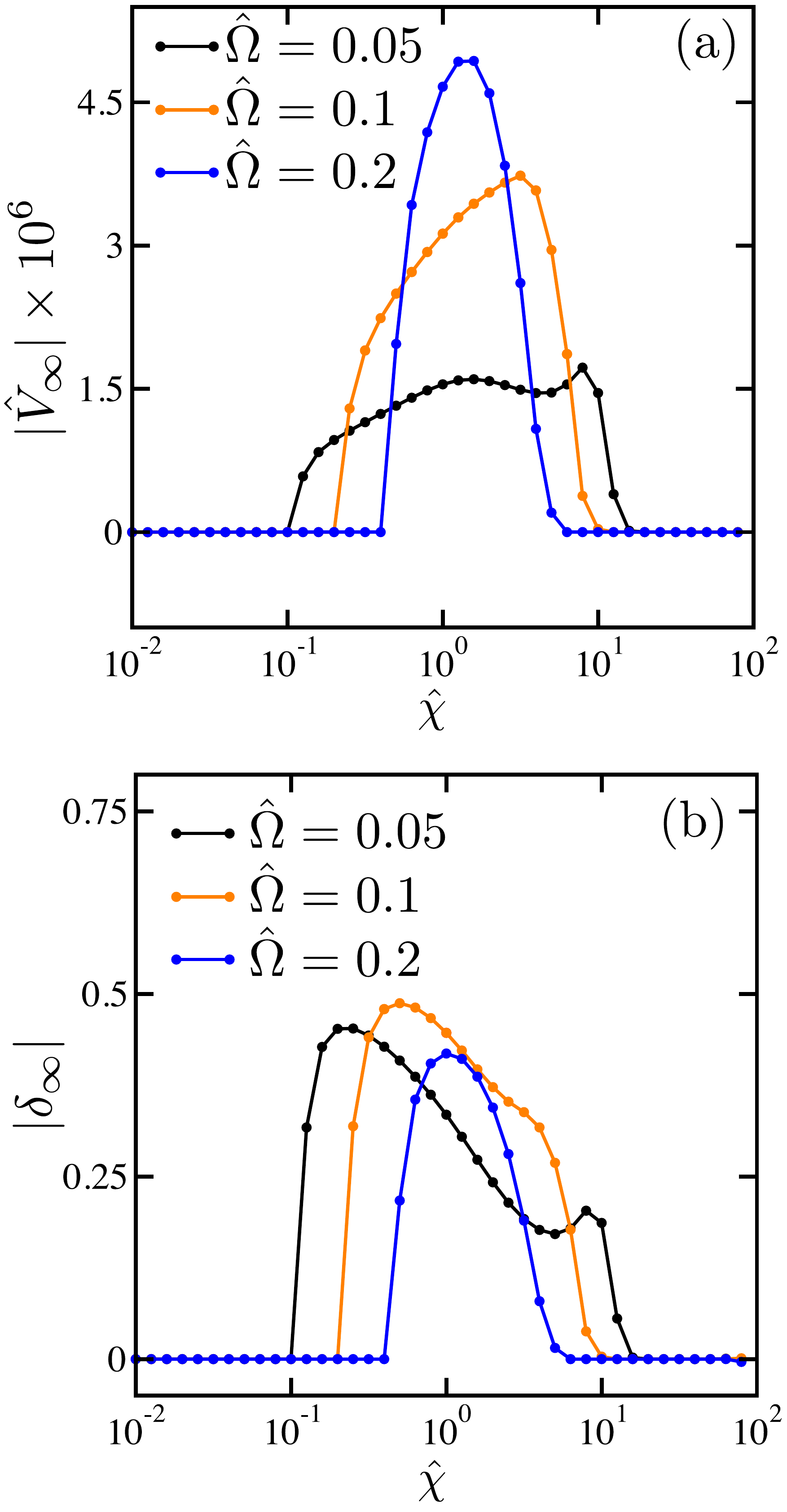}
\caption{
(Color online) 
Plots of (a) dimensionless stationary velocity $\vert \hat{V}_\infty \vert 
= \vert V_\infty \tau / \ell \vert$ and (b) stationary phase difference 
$\vert \delta_\infty \vert$ as a function of the dimensionless coupling parameter $\hat{\chi}=\chi \tau$
($\tau=6\pi\eta a/K_{\rm A}$).
In both plots, the dimensionless frequency is chosen as $\hat{\Omega}=0.05$ (black), $0.1$  (orange), 
and $0.2$ (blue), while $\delta_0=-\pi/2$ is fixed.
There is a lower critical value  $\chi_{\rm c}$ above which both  $\vert V_\infty \vert$ and 
$\vert \delta_{\infty} \vert$ become nonzero. 
$\vert V_\infty \vert$ and $\vert \delta_\infty \vert$ take maximum values 
at $\chi_{\rm m}>\chi_{\rm c}$, and they vanish for large $\chi$.
Reprinted from Ref.~\cite{Era21}.
$\copyright$ 2021 Institute of Physics.
}
\label{fig5}
\end{figure}

The physical meaning of Eqs.~(\ref{time_evolution_phase1}) and (\ref{time_evolution_phase2})
is that the phase $\theta_{\rm A}$ ($\theta_{\rm B}$) for the natural length and the mechanical phase 
$\phi_{\rm A}$ ($\phi_{\rm B}$) tend to be different due to the coupling term with $\chi$, as 
schematically explained in Fig.~\ref{fig4}.
Since the middle sphere is connected to the other two spheres, this model contains a feedback mechanism 
that regulates the dynamics of the two natural lengths $\ell_{\rm A}$ and $\ell_{\rm B}$.
Such a coupling effect in the spring motion gives rise to a non-reciprocal body motion and results in the 
autonomous locomotion of a microswimmer.

Let us define the time-dependent phase difference between the oscillations in the natural 
lengths by  
$\delta (t)=\theta_{\rm B}(t)-\theta_{\rm A}(t)$.
When $\chi=0$, the present model reduces to the elastic three-sphere microswimmer 
discussed in Sect.~\ref{sec:elastic}.
In this limit, we have $\theta_{\rm A}(t) = \Omega t$ and 
$\theta_{\rm B}(t) = \Omega t + \delta_0$, 
where $\delta_0=\delta(0)$ is the initial phase difference.
Hence, the initial phase difference $\delta_0$ and the frequency $\Omega$ fully 
determine the average velocity of locomotion when $\chi=0$.

When $\chi >0$, however, a stable phase difference $\delta$ controls the dynamics of a microswimmer 
irrespective of its initial value $\delta_0$.
Moreover, the transition to a non-reciprocal motion, as well as the average velocity, can be 
precisely tuned by $\chi$, and they are not solely fixed by the externally given frequency $\Omega$.
In Figs.~\ref{fig5}(a) and (b), we plot the numerically obtained steady-state velocity $\vert V_\infty \vert$
and the phase difference $\vert \delta_\infty \vert$, respectively, as a function of $\hat{\chi}=\chi \tau$ 
for different frequencies $\hat{\Omega}= \Omega \tau$ ($\tau=6\pi\eta a/K_{\rm A}$).
When $\hat{\Omega}=0.1$ (orange), for example, there is a critical value $\hat{\chi}_{\rm c} 
\approx 0.2 $ above which $\vert V_\infty \vert$ and $\vert \delta_\infty \vert$ become nonzero.
For $\hat{\chi} < \hat{\chi}_{\rm c}$, on the other hand, both $\vert V_\infty \vert$ and 
$\vert \delta_\infty \vert$ vanish. 
The existence of such a finite critical value $\hat{\chi}_{\rm c}$ is a nontrivial outcome of 
the present model.
When $\hat{\chi}$ is very large, such as $\hat{\chi}\ge 12.5$ for $\hat{\Omega}=0.1$,
both $\vert V_\infty \vert$ and $\vert \delta_\infty \vert$ vanish again. 
Hence autonomous locomotion can be achieved for a finite range of the coupling parameter $\chi$. 
Such behavior can also be observed for other frequencies $\hat{\Omega}$.

\section{Two Interacting Three-Sphere Microswimmers}
\label{sec:two}

Next, we discuss the behaviors of two interacting elastic three-sphere microswimmers in 
a viscous fluid as shown in Fig.~\ref{fig1}(e)~\cite{Kuroda19}.
The positions of the three spheres in the left (L) swimmer are denoted by $x_1$, $x_2$, and $x_3$,
while those in the right (R) swimmer are denoted by $x_4$, $x_5$, and $x_6$. 
We consider the situation when $x_1<x_2<x_3 \ll x_4<x_5<x_6$ is satisfied.
The distance between the two swimmers is defined by the positions of the two middle spheres, i.e.,
$\Delta=x_5-x_2$.

The equations of motion of each sphere ($i=1, \dots, 6$) are given by Eq.~(\ref{motion})
as before but we do not consider any noise here. 
We require two force-free conditions for each swimmer, i.e., 
$f_1+f_2+f_3=0$ and $f_4+f_5+f_6=0$.
Similar to Eq.~(\ref{armexpnasion}), we define the four displacements of the springs with 
respect to the natural length $\ell$ for the left and right swimmers as 
\begin{align}
& u^{\rm L}_{\rm A}(t)=x_2(t)-x_1(t)-\ell, \quad u^{\rm L}_{\rm B}(t)=x_3(t)-x_2(t)-\ell,
\label{uL} \\
& u^{\rm R}_{\rm A}(t)=x_5(t)-x_4(t)-\ell, \quad u^{\rm R}_{\rm B}(t)=x_6(t)-x_5(t)-\ell.
\label{uR}
\end{align}
The average velocities of the left and right swimmers can be calculated by 
$V^{\rm L}=( \dot{x}_1+\dot{x}_2+\dot{x}_3 )/3$ and 
$V^{\rm R}=( \dot{x}_4+\dot{x}_5+\dot{x}_6 )/3$,
respectively.

Under the condition that the two swimmers are far from each other and the deformations are small 
compared with $\ell$, i.e., $a \ll u^{\rm L,R}_{\rm A,B} \ll \ell \ll \Delta$,
one can perform a perturbative calculation to obtain the average velocities as
\begin{align}
& \overline{V}^{\rm L} = \frac{7a}{24\ell^2} \overline{
\left( u^{\rm L}_{\rm A} \dot{u}^{\rm L}_{\rm B}-u^{\rm L}_{\rm B} \dot{u}^{\rm L}_{\rm A} \right) }
\nonumber \\
& - \frac{a\ell }{\Delta^3} \overline{\left( 
u^{\rm R}_{\rm A}\dot{u}^{\rm R}_{\rm B}
- u^{\rm R}_{\rm B}\dot{u}^{\rm R}_{\rm A}
-u^{\rm L}_{\rm A}\dot{u}^{\rm R}_{\rm A} 
-u^{\rm L}_{\rm A}\dot{u}^{\rm R}_{\rm B}
+u^{\rm L}_{\rm B}\dot{u}^{\rm R}_{\rm A}
+u^{\rm L}_{\rm B}\dot{u}^{\rm R}_{\rm B}
\right)},
\label{lvellr}\\
& \overline{V}^{\rm R} = \frac{7a}{24\ell^2} \overline{
\left( u^{\rm R}_{\rm A} \dot{u}^{\rm R}_{\rm B}-u^{\rm R}_{\rm B} \dot{u}^{\rm R}_{\rm A} \right) }
\nonumber \\
& - \frac{a\ell}{\Delta^3} \overline{\left(
u^{\rm L}_{\rm A}\dot{u}^{\rm L}_{\rm B}
-u^{\rm L}_{\rm B}\dot{u}^{\rm L}_{\rm A}
-u^{\rm R}_{\rm A}\dot{u}^{\rm L}_{\rm A} 
-u^{\rm R}_{\rm A}\dot{u}^{\rm L}_{\rm B}
+u^{\rm R}_{\rm B}\dot{u}^{\rm L}_{\rm A}
+u^{\rm R}_{\rm B}\dot{u}^{\rm L}_{\rm B}
\right)}.
\label{rvellr}
\end{align}
Here we have kept only up to second-order terms in $u^{\rm L,R}_{\rm A,B}$, because of the condition
$u^{\rm L,R}_{\rm A,B}/\ell \ll \ell/\Delta$~\cite{Najafi10}.
The first terms on the right-hand side of the above equations represent the average swimming velocity 
of a single three-sphere swimmer, as we have obtained in Eq.~(\ref{velocitygeneral}).

The second terms on the right-hand side of Eqs.~(\ref{lvellr}) and (\ref{rvellr}) are due to the 
hydrodynamic interaction between the two swimmers.
These correction terms decay as $(\ell/\Delta)^3$ with increasing distance because they result from force 
quadrupoles rather than force dipoles.
The correction terms $(u^{\rm R}_{\rm A}\dot{u}^{\rm R}_{\rm B} 
- u^{\rm R}_{\rm B}\dot{u}^{\rm R}_{\rm A})$ in $V^{\rm L}$
and $(u^{\rm L}_{\rm A}\dot{u}^{\rm L}_{\rm B} 
-u^{\rm L}_{\rm B}\dot{u}^{\rm L}_{\rm A})$ in  $V^{\rm R}$ are both passive 
terms because they correspond to the swimming of only the second swimmer.
The other correction terms are due to the simultaneous motion of the two swimmers and 
hence are called active terms~\cite{Pooley07,Farzin12}.

More detailed analysis of Eqs.~(\ref{lvellr}) and (\ref{rvellr}) reveals that the mean of the two 
average velocities $(\overline{V}^{\rm L}+\overline{V}^{\rm R})/2$ is always smaller than 
that of a single elastic swimmer~\cite{Kuroda19}.
On the other hand, the velocity difference $\overline{V}^{\rm L}-\overline{V}^{\rm R}$ depends 
on the relative phase difference between the two elastic swimmers.
As a result, the swimming state of two elastic swimmers can be either bound or unbound 
depending on the relative phase difference~\cite{Kuroda19}. 
A more extended study on the interaction between two elastic microswimmers is given 
in Ref.~\cite{Ziegler21}.

\section{Three-Sphere Microswimmer in a Viscoelastic Fluid}
\label{sec:viscoelastic}

For microswimmers moving in soft materials, the surrounding fluid is not necessarily purely viscous 
but viscoelastic.
Several studies have discussed the swimming behaviors of microswimmers in different types of viscoelastic 
fluids~\cite{Lauga09,Teran10,Curtis13,Qiu14,Ishimoto17,Li21,Spagnolie23,Qin23}.
In this section, we discuss a three-sphere microswimmer in a general viscoelastic 
medium~\cite{Yasuda17b,Yasuda20}.
As shown in Fig.~\ref{fig1}(f), we consider the original three-sphere microswimmer as in 
Sect.~\ref{sec:original}.

The equation that describes the hydrodynamics of a low-Reynolds-number flow in  
a viscoelastic medium is given by the following generalized Stokes equation~\cite{Granek11}:
\begin{equation}
\int_{-\infty}^t dt'\,\eta(t-t') \nabla^2 \mathbf{v}(\mathbf{r},t') -\nabla P(\mathbf{r},t)=0.
\label{StorksEve}
\end{equation}
Here $\eta(t)$ is the time-dependent shear viscosity, $\mathbf{v}$ is the velocity field, 
$P$ is the pressure, and $\mathbf{r}$ stands for a three-dimensional positional vector.
The above equation is further subjected to the incompressibility condition, 
$\nabla\cdot\mathbf{v}=0$.

In the context of microrheology~\cite{MW95,GSOMS,FurstBook}, one uses a linear 
relation between the time-dependent force $F(t)$ acting on a sphere of radius $a$ 
and its time-dependent velocity $V(t)$ in the Fourier domain.
Such a linear response is written as $V(\omega)=\mu[\omega]F(\omega)$, 
where $V(\omega) =\int_{-\infty}^{\infty} dt \, V(t) e^{-{\mathrm i}\omega t}$, 
for example,
and the frequency-dependent self-mobility is given by $\mu[\omega] =(6\pi\eta[\omega] a)^{-1}$, 
where $\eta[\omega]=\int_{0}^{\infty} dt\, \eta(t)e^{-{\mathrm i}\omega t}$~\cite{GSOMS}. 
Similarly, the force $F_{j}(t)$ acting on the $j$-th sphere at $x_j$ and the induced velocity $V_i(t)$ 
of the $i$-th sphere at $x_i$ ($i \neq j$) are related by $V_i(\omega)=M[\omega]F_{j}(\omega)$, 
where $M[\omega]=(4\pi\eta[\omega]r)^{-1}$ is the frequency-dependent  
coupling mobility and $r=|x_i-x_j|$~\cite{Crocker00}.
By using these relations, the equations of motion for a three-sphere microswimmer in a 
viscoelastic fluid can be written similarly to Eq.~(\ref{motion}) in the Fourier domain.

We assume Eq.~(\ref{harmonicdeformation}) for the prescribed motion of the two arms. 
Up to the lowest order terms in $a$, the average swimming velocity over one 
cycle of motion is obtained as~\cite{Yasuda17b,Yasuda20}
\begin{equation}
\overline{V} = \frac{7a d_\mathrm Ad_\mathrm B \Omega}{24\ell^2}\frac{\eta'[\Omega]}{\eta_0}
\sin\phi 
-\frac{5a (d_\mathrm A^2-d_\mathrm B^2) \Omega}{48\ell^2}\frac{\eta''[\Omega]}{\eta_0},
\label{barVve}
\end{equation}
where $\eta'[\Omega]$ and $\eta''[\Omega]$ are the real and imaginary parts of the 
complex shear viscosity, respectively, and $\eta_0=\eta[\Omega \rightarrow 0]$ is the 
constant zero-frequency viscosity.
The first term can be regarded as the viscous contribution because it includes the real part 
$\eta'[\Omega]$, and is present only if the arm motion is non-reciprocal, i.e., $\phi \ne 0,\pi$.
The second term, on the other hand, corresponds to the elastic contribution because it 
contains the imaginary part $\eta''[\Omega]$, and it exists only when the structural symmetry 
of the swimmer is broken, i.e., $d_\mathrm A \ne d_\mathrm B$. 
Due to the presence of the second term, Purcell's scallop theorem can be generalized for 
viscoelastic fluids.
For a purely Newtonian fluid ($\eta[\Omega]=\eta_0$), the second term vanishes, and the 
first term coincides with Eq.~(\ref{velocityharmonic}).

\begin{figure}[t]
\begin{center}
\includegraphics[scale=0.5]{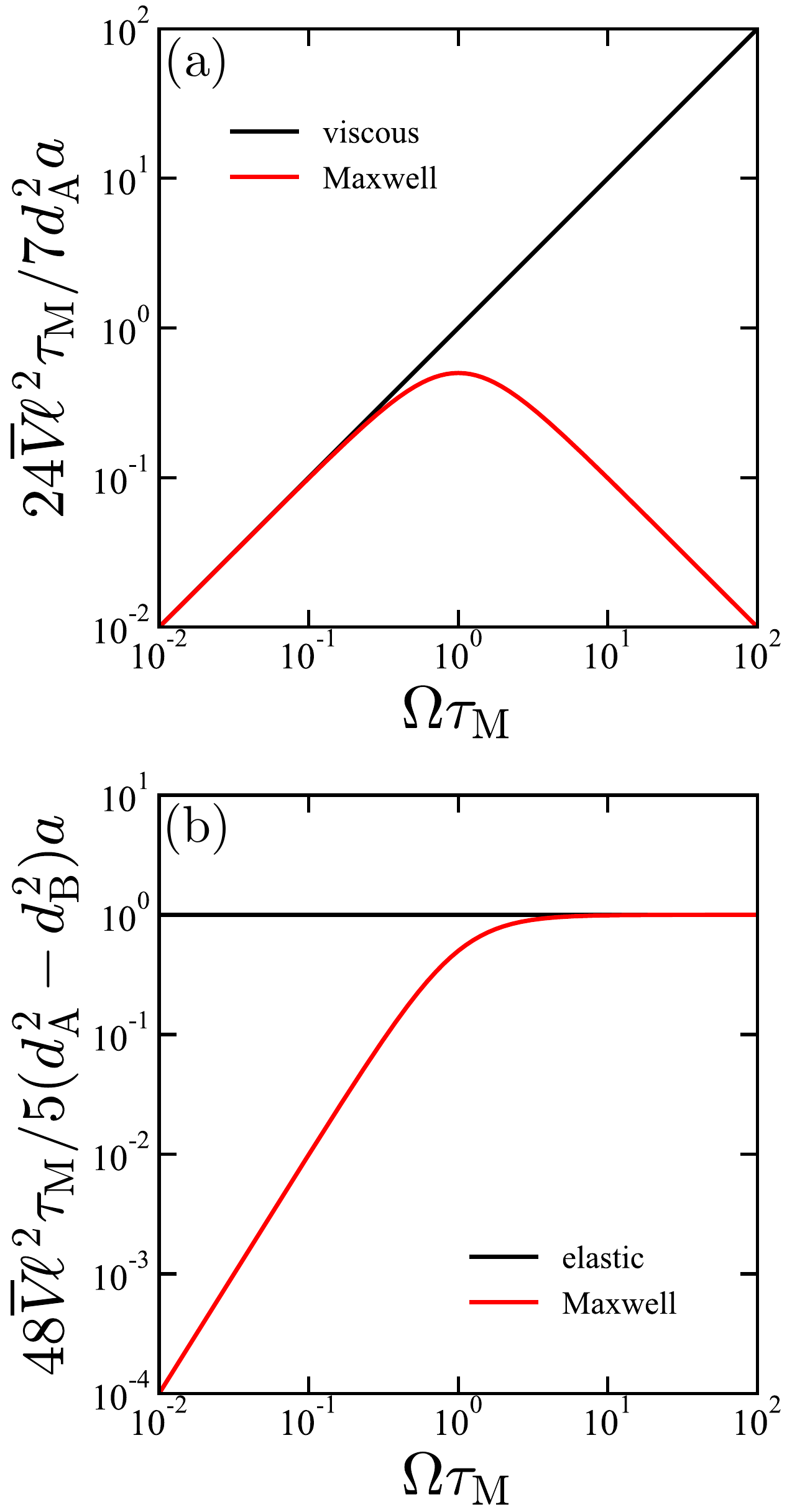}
\end{center}
\caption{
(Color online) Average swimming velocity $\overline{V}$ as a function of $\Omega \tau_{\rm M}$
for a three-sphere microswimmer in a viscoelastic Maxwell fluid [see Eq.~(\ref{maxwell})].
(a) Viscous contribution [the first term in Eq.~(\ref{maxwellV})] by setting $\phi=\pi/2$ and 
$d_{\rm A}=d_{\rm B}$. 
The case for a viscous fluid is plotted by the black line.
(b) Elastic contribution [the second term in Eq.~(\ref{maxwellV})] by setting $\phi=0$ and 
$d_{\rm A} \ne d_{\rm B}$.
The case for an elastic medium is plotted by the black line.
Reprinted from Ref.~\cite{Yasuda17b}.
$\copyright$ 2017 The Physical Society of Japan.
}
\label{fig6}
\end{figure}

As an illustration of the above result, we assume that the surrounding viscoelastic medium is 
described by the Maxwell model.  
In this case, the frequency-dependent viscosity can be written as 
\begin{equation}
\eta[\omega]=\eta_0\frac{1-{\mathrm i}\omega\tau_{\rm M}}{1+\omega^2\tau_{\rm M}^2}, 
\label{maxwell}
\end{equation}
where $\tau_{\rm M}$ is the characteristic time scale.
Such a medium is viscous for $\omega\tau_{\rm M} \ll 1$, while it becomes elastic for 
$\omega\tau_{\rm M} \gg 1$.
Then, the average swimming velocity in Eq.~(\ref{barVve}) becomes
\begin{align}
\overline{V} =\frac{7ad_\mathrm Ad_\mathrm B\Omega}{24\ell^2}\frac{1}{1+\Omega^2\tau_{\rm M}^2}
\sin \phi 
+\frac{5a(d_\mathrm A^2-d_\mathrm B^2)\Omega}{48\ell^2}
\frac{\Omega\tau_{\rm M}}{1+\Omega^2\tau_{\rm M}^2}.
\label{maxwellV}
\end{align}
The first viscous term increases as $\overline{V} \sim \Omega$ for $\Omega \tau_{\rm M} \ll 1$,
while it decreases as $\overline{V} \sim \Omega^{-1}$ for $\Omega \tau_{\rm M} \gg 1$.
On the other hand, the second elastic term increases as 
$\overline{V} \sim \Omega^2$ for $\Omega \tau_{\rm M} \ll 1$, and it approaches a constant 
value for $\Omega \tau_{\rm M} \gg 1$.
In Figs.~\ref{fig6}(a) and (b), we plot these viscous and elastic contributions to the average 
velocity $\overline{V}$, respectively, as a function of $\Omega \tau_{\rm M}$.
We note here the similarity between Figs.~\ref{fig2}(a) and \ref{fig6}(a) showing the decrease 
of the velocity in the high-frequency regime.

In the rest of this section, we mention two other situations; (i) a three-sphere 
microswimmer [see Fig.~\ref{fig1}(f)] and (ii) a two-sphere microswimmer [see Fig.~\ref{fig1}(g)], in which 
the microswimmer undergoes a reciprocal (rather than non-reciprocal) motion in a viscoelastic fluid~\cite{Yasuda20}. 
In the first case, we consider a three-sphere microswimmer whose two arms are subjected to 
different frequencies.
In particular, we consider the following time dependencies of the two arms:
\begin{align}
L_\mathrm A(t)=\ell+d\cos(\Omega t), \quad 
L_\mathrm B(t)=\ell+d\cos(2\Omega t),
\label{VEeq:armlength}
\end{align}
where the frequencies of $L_\mathrm A$ and $L_\mathrm B$ are $\Omega$ and $2\Omega$, 
respectively, while the amplitude $d$ is taken to be the same. 
(In general, the frequency of $L_{\rm B}$ can be $n \Omega$ where $n$ is an integer.)
Since the arm frequencies are different, a phase shift does not play any role, and 
the overall arm motion turns out to be reciprocal.
Nevertheless, the average velocity is obtained as 
\begin{align}
\overline{V} = \frac{5a d^2 \Omega}{48 \ell^2 \eta_0}
\left( 2 \eta''[2\Omega] - \eta''[\Omega] \right),
\label{VEeq:res1}
\end{align}
where only the imaginary parts of the complex shear viscosity appear.
The above result indicates that a reciprocal microswimmer can move as long as 
$\eta''[\Omega]\neq 2\eta''[2\Omega]$.

In the second case, we consider a two-sphere microswimmer consisting of two spheres having 
different sizes $a_1$ and $a_2$, as shown in Fig.~\ref{fig1}(g)~\cite{Yasuda20}.
These two spheres are connected by a single arm which undergoes the periodic 
motion $L(t)=\ell+d\cos{(\Omega t)}$.
Since there is only one arm, it is obvious that any periodic arm motion is inevitably reciprocal.
Calculating the total swimming velocity by $V=(\dot{x}_1+\dot{x}_2)/2$, we obtain its average as 
\begin{align}
\overline{V} = \frac{3 a_{1}a_{2}(a_{1}-a_{2})d^{2}\Omega}{4\ell^2(a_{1}+a_{2})^{2}\eta_0}
\eta''[\Omega].
\label{VEeq:res4}
\end{align}
This result shows that a reciprocal two-sphere microswimmer can swim in a viscoelastic fluid 
when the sphere sizes are different, i.e., $a_{1}\neq a_{2}$.
Similar to the first case in Eq.~(\ref{VEeq:res1}), $\overline{V}$ depends only on $\eta''[\Omega]$ 
representing the elastic contribution. 
These two examples further confirm that the scallop theorem can be generalized for viscoelastic 
fluids because various reciprocal deformations of a microswimmer can still induce its locomotion.

\section{Three-Sphere Microswimmer in a Structured Fluid}
\label{sec:structured}

The locomotion of a microswimmer discussed in the previous section is valid for 
homogeneous viscoelastic fluids without any internal structures. 
However, one of the characteristic features of soft matter is that it contains various intermediate 
mesoscopic structures and behaves as structured fluids~\cite{WittenBook}.
The existence of such internal length scales significantly affects the rheological properties of 
structured fluids~\cite{LarsonBook}.
In this section, we address the effects of intermediate structures of the surrounding viscoelastic 
fluid on the locomotion of a three-sphere microswimmer as shown in Fig.~\ref{fig1}(h)~\cite{Yasuda18}. 
Because a three-sphere microswimmer is characterized by its own size (the sphere size $a$ and 
the average arm length $\ell$), the  swimming velocity depends on the relative magnitudes of the 
swimmer's size and the characteristic length of the surrounding fluid.

Consider a viscoelastic structured fluid with a characteristic length scale $\kappa^{-1}$ and 
a time scale $\tau_{\rm v}$. 
We assume that the frequency-dependent one-point and two-point mobilities are expressed by 
the following scaling forms:
\begin{align}
\mu[a,\omega]=\frac{\hat \mu[\kappa a,\omega\tau_{\rm v}]}{6\pi\eta_0a}, \quad
M[r,\omega]=\frac{\hat M[\kappa r,\omega\tau_{\rm v}]}{4\pi\eta_0\ell},
\label{scaling}
\end{align}
where $\hat \mu$ and $\hat M$ are the dimensionless scaling functions and $\eta_0$ 
is the zero-frequency shear viscosity as before.
Unlike the homogeneous viscoelastic fluid, the mobilities $\mu$ and $M$ for a structured fluid 
are assumed to be written by the scaling functions that depend on the combinations 
$\kappa a$ and $\kappa r$.
Using these scaling functions, we write down the equation of motion for a three-sphere microswimmer 
in a structured fluid.
Then, the swimming velocity can generally be obtained in terms of $\hat\mu$ and $\hat M$, 
and we obtain both the viscous and elastic contributions as in the previous section~\cite{Yasuda18}.

To illustrate the importance of intermediate length scales in structured fluids, we consider a polymer 
gel described by the two-fluid model~\cite{deGennes76a,deGennes76b}.
There are two dynamical fields in this model: the displacement field $\mathbf{u}$ of the elastic network 
and the velocity field $\mathbf{v}$ of the permeating fluid [see Fig.~\ref{fig1}(h)].
When inertial effects are neglected, the linearized coupled equations for a polymer gel are given by 
\begin{align}
G \nabla^2 \mathbf{u}+\left(K+\frac{G}{3}\right)\nabla(\nabla\cdot \mathbf{u})-\Lambda 
\left(\frac{\partial \mathbf{u}}{\partial t} -\mathbf{v} \right)=0,
\label{TF-model1}
\end{align}
\begin{align}
\eta \nabla^2 \mathbf{v} -\nabla P 
-\Lambda \left(\mathbf{v}- \frac{\partial \mathbf{u}}{\partial t}\right)+\mathbf{f}=0,
\label{TF-model2}
\end{align}
where $G$ and $K$ are the shear and compression moduli of the elastic network, respectively,
$\eta$ is the shear viscosity of the fluid, $P$ is the pressure, and $\mathbf{f}$ is the external 
force density acting on the fluid component.
The elastic and fluid components are coupled via the mutual friction terms characterized 
by the friction coefficient $\Lambda$. 
When the volume fraction of the elastic component is small, we further require the incompressibility
condition, $\nabla \cdot \mathbf{v} = 0$.
The above two-fluid model contains the characteristic length $\kappa^{-1}=(\eta/\Lambda)^{1/2}$ 
and the characteristic time $\tau_{\rm v}=\eta/G$.
The former length scale roughly corresponds to the mesh size of a polymer network, and the latter 
time scale sets the viscoelastic relaxation time.

Diamant calculated the self-mobility of a sphere in a two-fluid gel under ``a sticking fluid and a 
free network" boundary condition at the surface of the sphere~\cite{Diamant15}.
On the other hand, the current authors previously obtained a general expression for the coupling mobility 
connecting the velocity $\mathbf{v}$ and the force $\mathbf{f}$ in the two-fluid model~\cite{YOKM17,YOK17}.
Using these results, we calculated the swimming velocity of a three-sphere microswimmer in a 
two-fluid gel as the sum of the viscous and elastic contributions~\cite{Yasuda18}.
Because the surrounding gel is characterized by the network mesh size, $\kappa^{-1}$, the following 
three different situations can be distinguished under the condition $a \ll \ell$:
(i) a large swimmer when $\kappa a\gg 1$ and $\kappa \ell \gg 1$, 
(ii) a medium swimmer when $\kappa a\ll 1$ and $\kappa \ell\gg 1$, and
(iii) a small swimmer when $\kappa a\ll 1$ and $\kappa \ell\ll 1$.

In the following, we briefly discuss the case of a large swimmer~\cite{Yasuda18}.
The behavior of the viscous contribution for a large swimmer exhibits a non-monotonic dependence 
on the frequency $\Omega$ [see Eq.~(\ref{harmonicdeformation})]. 
A careful analysis reveals that it behaves as 
$\Omega \to \Omega^{-1/2} \to \Omega \to \Omega^{-1} \to \Omega$ 
as $\Omega$ increases.
This non-monotonic behavior is more pronounced for larger sphere sizes. 
On the other hand, the frequency dependence of the elastic contribution crosses over as 
$\Omega^2\to\Omega^0$.
Several asymptotic expressions were also obtained.
For example, in the limit of $\Omega\tau_{\rm v}\to0$, the viscous contribution becomes 
\begin{align}
\overline{V} =\frac{31a^3 d_\mathrm Ad_\mathrm B\Omega}{144\ell^4}\sin\phi, 
\label{V1}
\end{align}
showing different dependence on $a$ and $\ell$ compared to Eq.~(\ref{velocityharmonic})
for a three-sphere microswimmer in a purely viscous fluid.

\section{Other Generalizations and Outlook}
\label{sec:discussion}

In this article, we have reviewed various extensions of the three-sphere microswimmer.
There are other extensions such as the case when one of the spheres has a larger 
radius~\cite{GolestanianEPJE2008,Nasouri19} or when the three spheres are 
arranged in a triangular configuration~\cite{Aguilar12}.
Montino and DeSimone considered the case, in which one arm is periodically actuated 
while the other is replaced by a passive elastic spring~\cite{Montino15}.
It was shown that such a microswimmer exhibits a delayed mechanical response of the 
passive spring with respect to the active arm.
Later, they analyzed the motion of a three-sphere swimmer with arms 
having active viscoelastic properties mimicking muscular contraction~\cite{Montino17}.
Later, Nasouri \textit{et al.}\ discussed the motion of an elastic two-sphere microswimmer, 
in which one of the spheres is a neo-Hookean solid~\cite{Nasouri17}.

Golestanian and Ajdari proposed a different type of stochastic microswimmer for which the 
configuration space of a swimmer generally consists of a finite number 
of distinct states~\cite{GolestanianAjdari08,Golestanian10}.
A similar idea was employed by Sakaue \textit{et al.} who discussed the propulsion of 
molecular machines or active proteins in the presence of hydrodynamic interactions~\cite{Sakaue10}. 
Later, Huang \textit{et al.}\ considered a modified three-sphere swimmer in a two-dimensional 
viscous fluid~\cite{Huang12}.
In their model, the spheres are connected by two springs, the lengths of which are assumed 
to depend on the discrete states that are cyclically switched.

A model of a three-disk microswimmer in a quasi-two-dimensional supported membrane has 
been discussed~\cite{Ota18}
Due to the presence of the hydrodynamic screening length in the quasi-two-dimensional 
fluid~\cite{Saffman75,Evans88}, the geometric factor appearing in the average velocity exhibits 
three different asymptotic behaviors depending on the microswimmer size and the screening length. 
This is in sharp contrast with a microswimmer in a three-dimensional bulk fluid that exhibits only a 
single scaling behavior. 
The swimming behaviors of a three-sphere microswimmer near a wall were also discussed~\cite{Daddi-Moussa-Ider18}.

The future extensions of the three-sphere microswimmer will involve combining it with other 
advanced technologies, such as nanotechnology, materials science, and artificial intelligence, 
to create a more sophisticated and versatile micro-robot~\cite{Tsang20,Hartl21,Zou22,Liu23}. 
These extensions could enable the microswimmer to perform even more complex tasks, such 
as targeted drug delivery to specific cells or tissues, or navigating through the human body 
to locate and repair damaged tissues.

\begin{acknowledgments}

We thank K.\ Era, K.\ Ishimoto, H.\ Kitahata, A.\ Kobayashi, Y.\ Koyano, L.-S.\ Lin, R.\ Okamoto,
and I.\ Sou for useful discussions and previous collaborations.
K.Y.\ acknowledges the support by Grant-in-Aid for JSPS Fellows (No.\ 22KJ1640) from the 
Japan Society for the Promotion of Science.
S.K.\ acknowledges the support by National Natural Science Foundation of China (Nos.\ 12274098 and 
12250710127) and the startup grant of Wenzhou Institute, University of Chinese Academy of Sciences 
(No.\ WIUCASQD2021041).
\end{acknowledgments}


\end{document}